\documentclass[useAMS,usenatbib]{mn2e}
\usepackage{graphicx,amssym}
\newcommand{\bc}{\begin{center}}
\newcommand{\ec}{\end{center}}
\usepackage{color}
\newcommand{\comment}[1]{}

\title[galaxy assembly bias]
      {Detection of galaxy assembly bias}

\author[Wang et al. ]
       { Lan Wang$^{1}$$\thanks{Email: wanglan@bao.ac.cn}$,
          Simone M. Weinmann$^{2}$,
          Gabriella De Lucia$^{3}$, 
          Xiaohu Yang$^{4,5}$
	\\
        $^1$Partner Group of the Max Planck Institute for Astrophysics, National Astronomical Observatories, \\
            ~~Chinese Academy of Sciences,  20A Datun Road, Chaoyang District, Beijing, China\\
	$^2$Leiden  Observatory, Leiden University, P.O. Box 9513, 2300 RA Leiden,  The Netherlands  \\
	$^3$INAF - Astronomical Observatory of Trieste, via G.B. Tiepolo 11, I-34143 Trieste, Italy\\ 
$^4$Center for Astronomy and Astrophysics, Shanghai Jiao Tong University, Shanghai 200240, China\\
        $^5$Shanghai Astronomical Observatory, Nandan Road 80, Shanghai
          200030, China} 
\begin{document}

\date{Accepted 2013 ???? ??. 
      Received 2013 ???? ??; 
      in original form 2013 ???? ??}

\pagerange{\pageref{firstpage}--\pageref{lastpage}} 
\pubyear{2013}

\maketitle

\label{firstpage}

\begin{abstract}
  Assembly bias describes the finding that the clustering of dark matter
  haloes depends on halo formation time at fixed halo mass. In this paper, we
  analyse the influence of assembly bias on galaxy clustering using both
  semi-analytical models (SAMs) and observational data.  At fixed stellar mass,
  SAMs predict that the clustering of {\it central} galaxies depends on the
  specific star formation rate (sSFR), with more passive galaxies having a
  higher clustering amplitude. We find similar trends using SDSS group
  catalogues, and verify that these are not affected by possible biases due to
  the group finding algorithm. Low mass central galaxies reside in narrow bins
  of halo mass, so the observed trends of higher clustering amplitude for
  galaxies with lower sSFR is not driven by variations of the parent halo
  mass. We argue that the clustering dependence on sSFR represent a direct
  detection of assembly bias. In addition, contrary to what expected based on
  clustering of dark matter haloes, we find that low-mass central galaxies in
  SAMs with larger host halo mass have a {\it lower} clustering amplitude than
  their counter-parts residing in lower mass haloes. This results from the fact
  that, at fixed stellar mass, assembly bias has a stronger influence on
  clustering than the dependence on the parent halo mass.
\end{abstract}


\section{Introduction}
\label{sec:intro}

Empirical models that link  galaxy properties statistically with their hosting
dark matter  (sub)halo masses include  the Halo Occupation  Distribution (HOD)
method  \citep{berlind2002, yang2003,  wang2006}, and  the  abundance matching
method \citep{vale2005, guo2010, moster2010}.   The HOD method places galaxies
into dark matter haloes by modeling the number and the spatial distribution of
galaxies in  haloes of  given mass.  The  abundance matching method  assumes a
one-to-one  correspondence between  galaxies and  subhaloes, with  higher mass
galaxies residing in  subhaloes of higher mass at the  time of accretion. Both
methods  assume that  galaxy  properties depend  only  on halo  mass, and  are
independent of any other properties of the parent halo. This assumption, which
also    applies   to    the   excursion    set    (Extended   Press-Schechter)
formalism  \citep{bond1991}, however, is  challenged by  the discovery  of the
assembly bias  effect found  by \citet{gao2005} for  haloes less  massive than
about $10^{13}h^{-1}M_{\odot}$.   Assembly bias indicates the  finding that at
the same halo  mass, earlier assembled haloes cluster more  than the ones that
assemble  late.   Evidence  that  galaxy  properties  might  depend  on  other
properties than  halo mass has also  been found in  different galaxy formation
models \citep{zhu2006, croton2007}.

\begin{figure*}
\bc
\hspace{-1.4cm}
\resizebox{14cm}{!}{\includegraphics{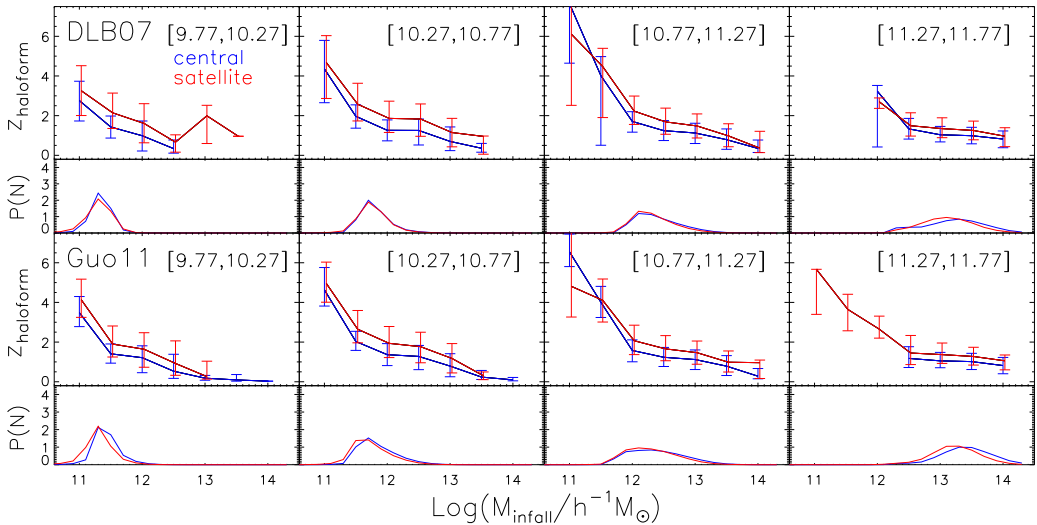}}\\%
\caption{ 
In the DLB07 (upper two panels) and the Guo11 (lower two panels) models,
the  halo formation  redshift  --  $M_{\rm infall}$  relation  and the 
fractional distribution (in unit of per Log($M_{\rm infall}$) dex)
in $M_{\rm  infall}$ in different stellar mass bins, 
for  central (blue lines) and  satellite (red lines)
galaxies.  Error bars  indicate the  68 percentile  distributions.  } 
\label{fig:Minfalltform}
\ec
\end{figure*}

From an observational point of view, \citet{yang2006} found that the
clustering of galaxy groups of fixed halo masses depends on the star
formation rate of  the central galaxies which are located at the centres of
dark matter haloes. This is probably  the first claim of
observational detection  of halo assembly  bias.  Along  this line,
subsequent  studies  have  also  searched  for a  correlation  between  galaxy
properties  and   halo  properties  at  fixed   halo  mass  \citep{wangyu2008,
tinker2011,  tinker2012}.  \citet{kauffmann2012}  have found  interesting
evidence that central galaxy properties  are correlated with the properties of
their    neighbours   beyond    the    virial   radius    of   their    parent
haloes \citep[``galactic  conformity''][]{weinmann2006}. The existence of 
assembly bias in the real Universe is, however, still debated (also because of  
uncertainties  in estimates of halo masses), and it remains
unclear  if  (and  to  what  extent) assembly  bias  influences  the  observed
properties of galaxies.
In this  paper, we analyse the  dependence of galaxy  clustering on different
galaxy  and halo  properties  in both  semi-analytic  galaxy formation  models
(SAMs) and in a group catalogue based on the SDSS DR7. Our main result is that
central galaxies with lower specific star formation rate cluster more than the
ones with  more active star formation at  fixed stellar mass, both  in SAM and
observations. As we  argue below, this may provide  evidence for assembly bias
in the real Universe.

The outline of our paper is as follows. In Section 2, we show the dependence
of galaxy clustering on halo mass, halo formation time, and specific star
formation rate of galaxies for two SAMs. In Section 3, observational
correlation functions for central galaxies split by specific star formation
rate and halo mass are calculated using group catalogues, and are compared
with the predictions from mock galaxy catalogues from the two SAMs we
analysed.  Conclusions and discussions are presented in Section 4. All model
results shown below are based on dark matter halo trees from the Millennium
Simulation \citep{springel2005}. The resolution of the simulation gives a 
lower limit in the dark matter halo mass of around $10^{11}h^{-1}M_{\odot}$, 
and a lower limit in the galaxy stellar mass of around $6\times10^{9} M_{\odot}$.

\section{effect of assembly bias in semi-analytic models}
\label{sec:SAM}

In our previous study \citep{wang2012b}, we have shown that the stellar masses
of galaxies extracted from  the semi-analytic models of \citet[][DLB07]{DLB07}
and  \citet[][Guo11]{guo11} depend  not only  on halo  mass but  also  on halo
formation  time.  In  this  section, we  investigate  further the  correlation
between galaxy  properties and those of  their parent haloes, with  the aim of
identifying likely indications of assembly bias. In the following, 
$M_{\rm  infall}$ is the  mass of  host  halo when  galaxy is/was  last time  a
central object  of its  hosting FOF  group. 
For centrals, $M_{\rm  infall}$=$M_{\rm  halo}$.

\begin{figure*}
\bc
\hspace{-0.4cm}
\resizebox{17cm}{!}{\includegraphics{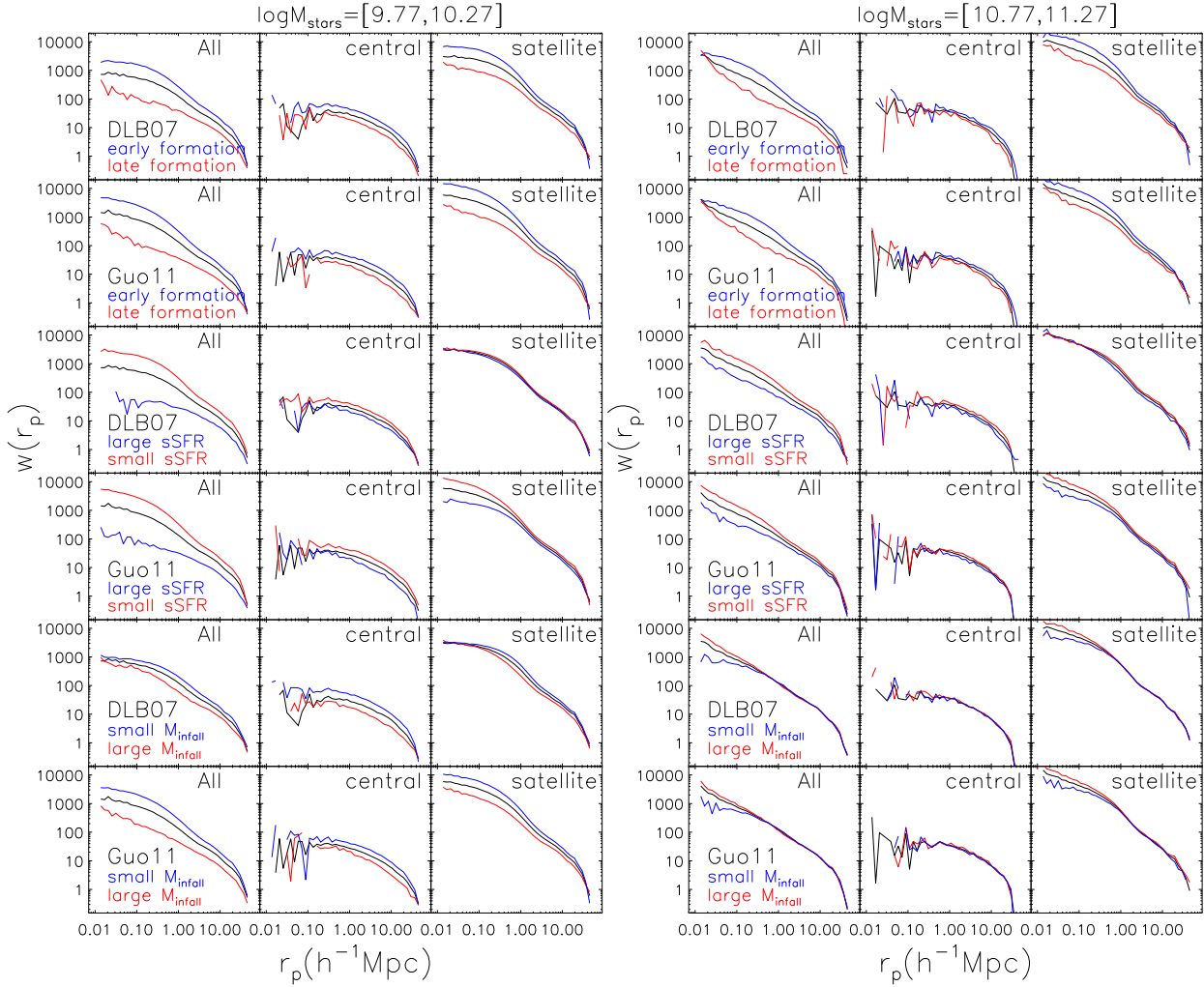}}\\%
\caption{ 
Correlation functions (CFs) of galaxies split by halo formation time, sSFR and
$M_{\rm   infall}$    in   two   stellar   mass   bins    of   $\log   (M_{\rm
star}/M_{\odot})$=[9.77,10.27] and  [10.77,11.27] in  the two SAMs.   For each
stellar mass bin, results are shown for  all galaxies in the mass bin, as well
as for central and satellite galaxies separately.  In each panel, a black line
is used to show  the CF for all galaxies in the  corresponding mass bin, while
coloured lines  correspond to sub-samples  with values of  the considered
physical property  smaller or larger than the  median. }
\label{fig:CFmore}
\ec
\end{figure*}

Fig.~\ref{fig:Minfalltform} shows  the relation between the  formation time of
host haloes (defined as the redshift when half of $M_{\rm   infall}$ 
was first assembled into a single object) and $M_{\rm  infall}$ in two  SAMs. 
For a satellite galaxy, we trace back its main progenitors to the time
when it was for the last time a central galaxy, and record the mass of its
parent FOF group at this time as $M_{\rm infall}$. To define the corresponding
formation time, we then follow the main progenitor of the halo identified at
the infall time, up to the time when its mass first reaches half of $M_{\rm
infall}$.  At all stellar masses, less massive haloes assemble much earlier
than more massive ones.  We also show in Fig.~\ref{fig:Minfalltform} the
distribution of $M_{\rm infall}$ for these galaxies. At low stellar masses, the
distributions are quite narrow.  We note that the bias varies very weakly as a
function of halo mass for haloes with $M_{\rm infall}\la
10^{12.5}h^{-1}M_{\odot}$
\citep[e.g.][]{Sheth2001}: for  the two lowest stellar mass bins
considered, the  maximum variation for the bias is  at the level of $10-15 \%$.
Therefore,  for  low stellar masses, large variations in the clustering as a 
function of galaxy physical properties cannot be attributed to  variations of
halo masses. For massive  galaxies, on  the  other hand, the distributions of 
halo masses  are rather  extended  and for haloes more  massive than 
$\sim 10^{12.5}h^{-1}M_{\odot}$ the  bias increases rapidly as a function of 
halo mass.  Therefore,  for  massive galaxies, one
expects a stronger clustering variation due to a wider range of host halo masses.

In Fig.~\ref{fig:CFmore}, we check the dependence of galaxy clustering on halo
formation time, sSFR, and $M_{\rm infall}$ at fixed stellar mass in the models
of DLB07 and Guo11 in two stellar mass bins. For each stellar mass bin,
galaxies are split into different sub-samples according to the median value of
each physical parameter considered. We show results both for the full
population of galaxies, and for centrals and satellites separately.  The top
two panels show that galaxies residing in earlier formed haloes cluster more
strongly than their counterparts residing in haloes that formed later, which
reflects the halo assembly bias \citep{gao2005, zhu2006, croton2007}.

When split by sSFR, we find that galaxies with lower sSFR cluster more. This is
consistent with observations that old and red galaxies cluster more even at
fixed stellar masses \citep{li2006a,bamford2009,weinmann2011} and is mainly due
to the higher clustering of satellites vs. centrals.  However, we find
that the result is still present when considering only central galaxies, or
only satellite galaxies at scales smaller than 1$h^{-1}$Mpc in the Guo11
model. In both SAMs, for the small stellar mass bin considered, the clustering
of central galaxies depends on their sSFR. We will show later that this trend
is not driven by trends as a function of parent halo mass.

\begin{figure*}
\bc
\hspace{-0.4cm}
\resizebox{14cm}{!}{\includegraphics{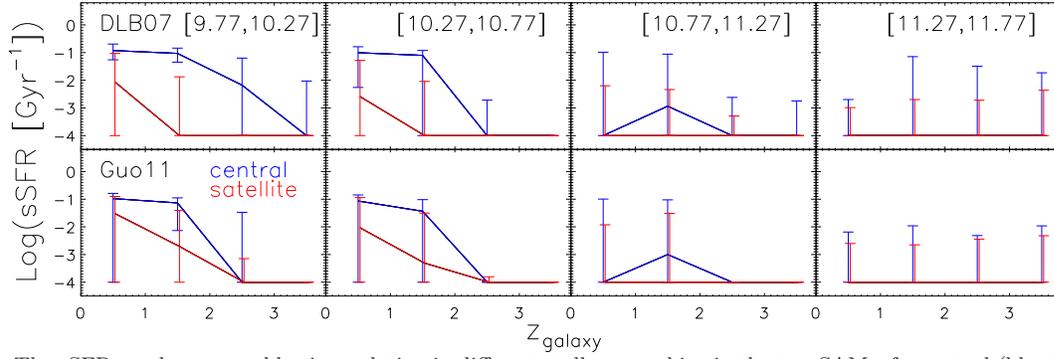}}\\%
\caption{ 
The sSFR -- galaxy assembly time relation in different stellar mass bins
in the two SAMs, for central (blue lines) and
satellite (red lines) galaxies. Error bars indicate 68 percentile
distribution. }
\label{fig:SSFRtform}
\ec
\end{figure*}

In Fig.~\ref{fig:SSFRtform}, we show the relation between galaxy sSFR and
galaxy assembly time, defined as the time when half of the galaxy stellar mass
of the present day is assembled in a single progenitor of the galaxy.  In
general, for galaxies with stellar mass lower than $10^{10.77}M_{\odot}$,
galaxies with lower sSFR assemble earlier. Combining this result with the
finding that low-mass central galaxies with lower sSFR cluster more, we find
that the clustering of low-mass central galaxies in SAMs depends on the galaxy
assembly time, i.e. there is an assembly bias.

The result that galaxies with low sSFR cluster more
than their counter-parts with higher sSFR could be due to a dependence of
galaxy clustering on halo mass: the clustering amplitude increases with
increasing halo mass, so if galaxies with lower sSFR are sitting in more
massive haloes, this would explain the trends found. However, this is
not the case for low mass galaxies. In the bottom two rows of
Fig.~\ref{fig:CFmore}, we show the galaxy correlation functions (CFs) split by
$M_{\rm infall}$. For low mass galaxies, those with \emph{smaller $M_{\rm
infall}$ actually cluster more strongly} in both SAMs, for both centrals and
satellites. This can be understood as follows: as shown in
Fig.~\ref{fig:Minfalltform}, at fixed stellar mass, less massive haloes form
earlier. And earlier formed haloes cluster more due to halo assembly bias. In
addition, the bias of the host haloes of low mass galaxies varies little as
mentioned before, which results into little variations of galaxy clustering
amplitude due to varying host halo mass. Therefore, for the lower stellar mass
bin considered, the dependence of clustering on halo formation time has a
larger influence than the dependence of clustering on halo mass. For the
higher stellar mass bin, the dependencies on halo mass and halo formation time
compensate, and CFs of galaxies with different $M_{\rm infall}$ are quite
similar, with only slight differences for satellites on small scales.

\section{assembly bias in observation}
\label{sec:obs}

Can we find assembly bias as seen in the SAMs also in reality? While it is
impossible to measure the halo formation time for observational results, we
can check galaxy clustering as a function of halo mass and sSFR, which could
be affected by assembly bias, as shown in Section 2.
 
For satellite galaxies, besides the possible effect of assembly bias on their
clustering property, there are other complications due to the fact that they
can be accreted over a range of cosmic epochs. Therefore, in the following, we
focus only on central galaxies.

\subsection{SDSS results}

We make use of the galaxy \citep{Blanton2005} and group
 \citep{yang2007} catalogues extracted from the SDSS DR7 data, and use these
catalogues to distinguish between central and satellite galaxies. For central
galaxies within different stellar mass bins, we split galaxies into two
subsamples with equal numbers of galaxies, according to the median value of
sSFR and $M_{\rm halo}$ respectively in this mass bin, and calculate the projected 
CFs for each
subsample.  The details of how to get the projected 2PCFs of SDSS DR7 galaxies
can be found in Appendix A of \citet{yang2012}.

In Fig.~\ref{fig:SDSSwrp}, CFs of subsamples in observation are shown for four
different stellar mass bins.  At low stellar masses, central galaxies with
lower sSFR cluster more than those with higher sSFR, which is consistent with
the trends found in the SAM. As argued in Section 2, this trend possibly
indicates a signature of galaxy assembly bias in the data.
Fig.~\ref{fig:SDSSwrp} also shows that there is no significant difference
between the clustering properties of sub-samples split by $M_{\rm halo}$ in the
group catalogues. Note, however, that in the SDSS observations (as well as in
the mock catalogues that we will use in the next subsection), halo masses are
estimated from a simple ranking of the group stellar masses. In this way,
splitting the sample as a function of halo mass just reflects trends as a
function of stellar mass. Any possible formation time information associated
with the mass of haloes in the real Universe will 
be erased due to the halo mass measurement adopted in the data.

\subsection{Comparison between SDSS and SAM mock catalogue}

\begin{figure*}
\bc
\hspace{-0.4cm}
\resizebox{14cm}{!}{\includegraphics{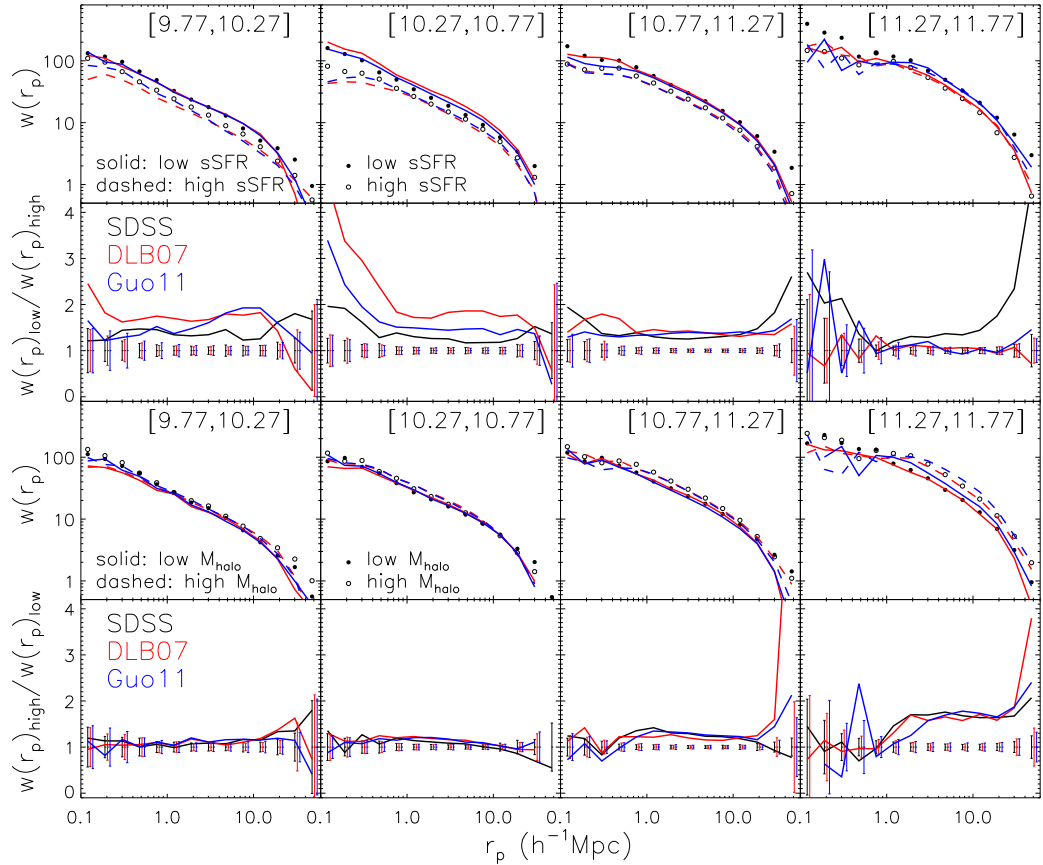}}\\%
\caption{ 
CFs of central galaxies split by sSFR (upper two rows) and $M_{\rm halo}$
(lower two rows) in the mock group catalogues of DLB07 and Guo11 (solid/dashed
lines), compared with
the results of SDSS group catalogue (filled/open circles), in four different stellar mass
bins. The ratios of CFs of subsamples with lower and higher sSFR (second row) and 
the ratios of CFs of subsamples with higher and lower $M_{\rm halo}$ (fourth row) are 
also presented, where error bars are the relative errors of the CFs for subsamples 
with higher sSFR and lower $M_{\rm halo}$, respectively. }
\label{fig:SDSSwrp}
\ec
\end{figure*}

To ensure a fair comparison with SAMs, we build mock catalogues for both DLB07
and Guo11 models, to mimic volumes and apparent magnitude limits of the
observational data and to take into account the observational selection
effects.  The mock redshift catalogues are constructed in a way similar to that
described in \citet{yang2004}, where the detailed sky coverage of the SDSS DR7
including the angular variations in the magnitude limits and completeness of
the data is taken into account \citep[see e.g.][]{li2007}. Then, we adopt the
same method used for the SDSS galaxies to define central and satellite
galaxies, and to compute their CFs. The results from these models are also
shown in Fig.~\ref{fig:SDSSwrp}. Similar to what shown in
Fig.~\ref{fig:CFmore}, for SAMs, central galaxies with lower sSFR cluster more
than the ones with higher sSFR. The trend is consistent with SDSS measurements.
Note that for central galaxies, the ratios between subsamples split in
sSFR in the Guo11 model are closer to the observational data, with respect to
predictions from the DLB07 model. When satellite galaxies are also taken
into account, however, the CFs in the Guo11 model over-predicts the
observational measurements for low-mass galaxies \citep{guo11, wang2012b}.

Comparing SAM mock catalogues and the results as presented in
Fig.~\ref{fig:CFmore}, we find some discrepancies. At small scales, the
clustering in mock catalogues is somewhat enhanced which might due to a small
fraction of mis-classified centrals. Nevertheless, for the low-mass bin, at
intermediate scales of $r_p$=[0.5, 20]$h^{-1}$Mpc, results are quite
consistent between the mock group catalogue and the original SAM catalogues,
and the dependence of clustering on sSFR is not affected by the algorithm used
to detect groups and identify central and satellite galaxies.  Thus, we
conclude that the finding that low sSFR centrals in the SDSS cluster more than
their more active counterparts of the same stellar mass is real, and not due
to spurious effects introduced by the group finding algorithm.

When splitting galaxies by host halo masses, results from the SDSS and those
from mock catalogues based on the SAMs used in our studies are consistent for
galaxies less massive than $10^{10.77}M_{\odot}$. However, the assembly
bias trends associated with $M_{\rm infall}$ shown in the lower-left panels of
Fig. ~\ref{fig:CFmore} do not show up in the mock catalogues. The reason is
already mentioned in previous subsection that in assigning halo masses to
groups, no additional information like formation time is kept.

\section{conclusion and discussions}
\label{sec:con}

In this paper, we find indications that the strong assembly bias
of low-mass galaxies in the semi-analytic galaxy formation models
of \citet {DLB07} and \citet{guo11} leaves a signature on galaxy clustering as
a function of halo mass and sSFR.  In particular, we find that central
galaxies with low sSFR galaxies cluster more than their counterparts with the
same stellar mass but higher sSFR.

A similar trend is found in group catalogues based on the SDSS DR7. For
low mass galaxies, this is likely a signature of assembly bias in the real
Universe.  Alternative explanations are possible: for example, the central
galaxies with lower sSFR could be associated with a `back-splash' population
of galaxies that have been satellites in the past. These galaxies are found to
form earlier than central galaxies that did not experience such an
event \citep[][Li et al., in preparation]{wang2009}, which may contribute
to assembly bias.  However, the fraction of central backsplash galaxies is
lower than 10 per cent, and therefore does not dominate the assembly bias
effect, as shown by \citet{wang2009}.

In SAMs, low mass central galaxies show a halo mass dependence 
that seems to contradict the halo model predictions: here, low mass 
haloes cluster stronger than high mass haloes. The reason is that, for these low 
mass central galaxies, there is a strong correlation between host
halo mass and halo formation time, with less massive haloes forming
earlier. Haloes that form early cluster more than haloes of similar mass that
form later (assembly bias). In addition, the clustering amplitude of the parent
haloes of low-mass central galaxies varies very little. Therefore, the
dependence of the clustering amplitude on halo formation time has a larger
impact than the dependence due to variations of parent halo mass.

More generally, we have found evidence that the clustering of central galaxies
does not only depend on host halo mass, but additionally on secondary
parameters like the specific star formation rate. While the physical reason 
for these trends remains unclear -- it may be solely assembly bias or a  
combination of different 
effects -- our results show that the simple picture that galaxy  properties 
depend only on halo mass is incomplete. This has to be taken into account in 
empirical models like HOD and abundance matching that  are based on this  
assumption, and for precision measurements of cosmological parameters using 
clustering studies.

\section*{Acknowledgments}

LW acknowledges support from the National basic research program of China (973
program under grant No. 2009CB24901), the NSFC grants program (No. 11103033, 
No. 11133003), and the Partner Group program of the Max Planck
Society.  SMW acknowledges funding from ERC grant HIGHZ no. 227749.  GDL
acknowledges financial support from the European Research Council under the
European Community's Seventh Framework Programme (FP7/2007-2013)/ERC grant
agreement n. 202781.

The simulation used in this paper was carried out as part of the programme of
the Virgo Consortium on the Regatta supercomputer of the Computing Centre of
the Max–Planck–Society in Garching.  The halo data, together with the galaxy
data from two semi-analytic galaxy formation models, are publicly available at
http://www.mpa-garching.mpg.de/millennium/.

\label{lastpage}

\bibliographystyle{mn2e}
\bibliography{scatter}

\end{document}